\documentclass[aps,prl,twocolumn,showpacs,10pt,nofootinbib]{revtex4-1}
\usepackage[dvips]{graphicx}
\usepackage{color,array,dcolumn}
\usepackage{amsmath}
\usepackage{amssymb}
\usepackage{mathrsfs}
\usepackage[mathscr]{euscript}

\begin{document}

\title{Spin Helix of Magnetic Impurities in Two-dimensional Helical
  Metal }

\author{Fei Ye\textsuperscript{1,2}, Guo-Hui Ding\textsuperscript{3},
  Hui Zhai\textsuperscript{2} and Zhao-Bin Su\textsuperscript{4}}

\affiliation{1. College of Material Science and Optoelectronics
  Technology, Graduated University of Chinese Academy of Science,
  Beijing 100049, P. R. China}

\affiliation{2. Institute for Advanced Study, Tsinghua University,
  Beijing 100084, P. R. China}

\affiliation{3. Department of Physics, Shanghai Jiaotong University,
  Shanghai 200240, P. R. China}

\affiliation{4. Institute of Theoretical Physics, Chinese Academy of
  Science, Beijing 100089, P. R. China}

\begin{abstract}
  We analyze the Ruderman-Kettel-Kasuya-Yosida(RKKY) interaction between
  magnetic impurities embedded in the helical metal on the surface of
  three-dimensional topological insulators. Apart from the conventional
  RKKY terms, the spin-momentum locking of conduction electrons also
  leads to a significant Dzyaloshinskii-Moriya (DM) interaction between
  impurity spins. For a chain of magnetic impurities, the DM term can
  result in single-handed spin helix on the surface. The handedness of
  spin helix is locked with the sign of Fermi velocity of the emergent
  Dirac fermions on the surface. We also show the polarization of
  impurity spins can be controlled via electric voltage for dilute
  magnetic impurity concentration.
\end{abstract}
\pacs{72.25.Dc, 73.20.-r, 75.30.Hx, 85.75.-d}
\maketitle

It is now known that a three-dimensional(3D) insulators with time
reversal symmetry can be classified into two categories: ordinary and
topological insulator(TI)
\cite{fu2007,moore2007,roy2006,qi2008,schnyder2008}.  Though both of
them are fully gapped in the bulk, TI has metallic surface states robust
against any time reversal invariant perturbations.  These metallic
surface states dubbed as ``\emph{helical metal}" can be described by a
two-dimensional(2D) massless Dirac equation. Comparing to other emergent
relativistic materials such as single-layer graphene, the surface states
of TIs have two unique features. One is that the number of Dirac nodes
is odd. In fact, according to a no-go theorem\cite{wu2006}, the Dirac
nodes always appear in pairs in the conventional 2D lattice such as
graphene. A surface state of TI can invalidate no-go theorem because a
pair of Dirac nodes are separated onto two opposite surfaces. TI with
single Dirac cone has been theoretically studied and experimentally
observed by angle-resolved photo-emission
spectroscopy\cite{Hsieh2008,Zhang2009,Xia2009,Chen2009,zhangyi2009}.
The other feature is that electron spin and momentum are intimately
locked in helical metal, originated from strong spin-orbit
coupling. Evidence of the spin-momentum locking has also been observed
in recent
experiments\cite{Hsieh2009,Roushan2009,zhangtong2009,alpichshev2010}.

\begin{figure}[htbp]
\includegraphics[width=4cm]{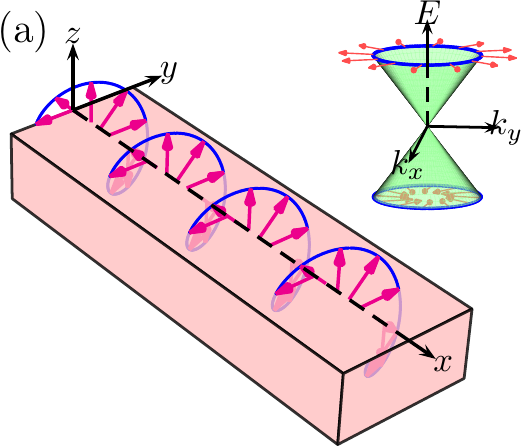}%
\includegraphics[width=4cm]{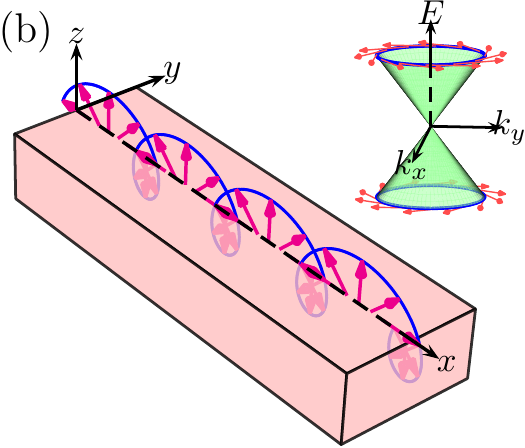}%
\caption[]{\label{fig:helix} Illustration of spin helix of a chain of
  magnetic impurities. Two types of Dirac Hamiltonian, $\hat{H}^a_0$ and
  $\hat{H}^b_0$(see \emph{Model} section), are considered, which lead to
  spin helix (magenta vectors) perpendicular(a) and parallel(b) to the
  chain direction, respectively. The insets show spin orientation of
  conduction electron in momentum space.}
\end{figure}

The effect of magnetic impurities in a metal is a central issue in
condensed matter physics, since the coupling between impurity and
conduction electrons can lead to many intriguing phenomena such as Kondo
effect and Ruderman-Kettel-Kasuya-Yosida(RKKY) interaction. The analogy
of these effect in a helical metal often reveals intrinsic property of a
TI, which has received considerable attention recently
\cite{feng2009,gao2009,liuQ2009,balatsky2009}. In this letter we point
out a novel feature in RKKY interaction between impurity spins in
helical metal, which manifests directly the two features of TI mentioned
above. Firstly, due to the spin-momentum locking of conduction electron,
we show the induced RKKY interaction must contain an anisotropic
Dzyaloshinskii-Moriya(DM) term by both symmetry analysis and
perturbation calculation, which is usually absent in systems like
graphene\cite{brey2007,saremi2007,bunder2009}. Since the emergent Dirac
Hamiltonian of helical metal is a pure spin-orbit coupling, the
magnitude of DM term is of the same order of the conventional RKKY one.
Secondly, the DM term can lead to \emph{single-handed} spin helix if the
impurities are aligned into a chain, of which the handedness depends on
the sign of the Fermi velocity $v_F$ of the helical metal.  For strong
TIs with a single Dirac cone attached to each surface, $v_F$ is fixed,
therefore the corresponding spin helix must be single-handed. The spin
helix is illustrated in Fig.~\ref{fig:helix}, where two types of Dirac
fermions are considered and the spin helix are
perpendicular(Fig.~\ref{fig:helix}a) and parallel
(Fig.~\ref{fig:helix}b) to impurity chain, respectively. This is a
hallmark of a strong TI, and could be possibly detected by
spin-polarized scan tunneling microscopy, or optical measurement like
Kerr effect.

\emph{Model}: We start from the Hamiltonian of the conduction electrons
with only one Dirac cone
\begin{eqnarray}
\label{eq:1}
\hat{H}^{a}_0=\sum_{\mathbf{k}}\hat{c}^{\dagger}_{\mathbf{k}} (\hbar
v_F\mathbf{k}\cdot\boldsymbol{\sigma}-E_F
\mathbf{1})\hat{c}_{\mathbf{k}},
\end{eqnarray}
with Fermi energy $E_F$ and Pauli matrix $\boldsymbol{\sigma}$.  Via a
unitary transformation $\hat{a}_{\mathbf{k}s} =[
\hat{c}_{\mathbf{k}\uparrow}+\text{sgn}(v_F)s e^{-i\theta_{\mathbf{k}}}
\hat{c}_{ \mathbf{k}\downarrow}]/\sqrt{2}$ with $\theta_{\mathbf{k}}$
the angle of vector $\mathbf{k}$ and $s=\pm1$, it can be diagonalized as
$\hat{H}^{a}_0 = \sum_{s=\pm1}(s\hbar
|v_F|k-E_f)\hat{a}^{\dagger}_{\mathbf{k}s} \hat{a}_{\mathbf{k}s}$. In
practice, the recently discovered TI material, e.g.,
$\text{Bi}_{2}\text{Se}_{3}$ has surface state described by the Rashba
type of Hamiltonian $\hat{H}^b_0= \sum_{\mathbf{k}}
\hat{c}^{\dagger}_{\mathbf{k}} [\hbar v_F (\mathbf{k}\times
\boldsymbol{\sigma})_{z} - E_f]
\hat{c}_{\mathbf{k}}$\cite{Zhang2009,Xia2009}, which is related to
$\hat{H}^a_0$ by a rotation of $\mathbf{k}$ around $z$ axis by $\pi/2$,
hence we can focus on $\hat{H}^a_0$ in the following.

Suppose there are $N_{imp}$ impurity spins located at $\mathbf{r}_n$
denoted by $\mathbf{S}_n$, which interact with conduction electrons via
the following spin-spin coupling
\begin{eqnarray}
\label{eq:2}
\hat{H}_I = \sum_{n=1}^{N_{imp}} \lambda_z
  \hat{s}^z(\mathbf{r}_n) \hat{S}^z_{n}+ \lambda_{\pm}
[\hat{s}^{x}(\mathbf{r}_n)\hat{S}^{x}_{n}+\hat{s}^{y}(\mathbf{r}_n)
\hat{S}^{y}_n],
\end{eqnarray}
where $\hat{s}^{x,y,z}(\mathbf{r})$ are the spin operators of the conduction
electron at $\mathbf{r}$. The coupling constants are assumed to be
isotropic in $xy$ plane $\lambda_x=\lambda_y\equiv\lambda_{\pm}$, but
may differ from the $z$-component. The total Hamiltonian is simply the
sum $\hat{H} = \hat{H}^{a,b}_0+ \hat{H}_I$.

Before proceeding with the discussion of the effective interaction
between impurity spins, we follow
RKKY\cite{ruderman1954,Kasuya1956,kei1957} to divide $\hat{H}_I$ into
two parts: $\hat{H}_{I0}$ and $\hat{H}_{I1}$. $\hat{H}_{I0}$ consists of
the diagonal term of
$\hat{a}^{\dagger}_{\mathbf{k}s}\hat{a}_{\mathbf{k}s}$ and can be
written as $\hat{H}_{I0}=\lambda_{\pm}\sum_{\mathbf{k}}
(\mathbf{e}_{\mathbf{k}}\cdot \hat{\mathbf{s}}_{\mathbf{k}})
\left(\mathbf{e}_{\mathbf{k}}\cdot \hat{\mathbf{S}}_{imp}\right) $,
where $\mathbf{e}_{\mathbf{k}}=\mathbf{k}/k$,
$\hat{\mathbf{s}}_{\mathbf{k}}
=\hat{c}^{\dagger}_{\mathbf{k}}\boldsymbol{\sigma}
\hat{c}_{\mathbf{k}}$, and $\hat{\mathbf{S}}_{imp}= \mathcal{V}^{-1}
\sum_{n=1}^{N_{imp}}\hat{\mathbf{S}}_n$ is the density of impurity
spin. And the remaining part is off-diagonal denoted by
$\hat{H}_{I1}$. Notice that for the helical Hamiltonian, the electric
current of the conduction electrons is proportional to their spins, for
instance, $\hat{\mathbf{J}} =ev_F
\sum_{\mathbf{k}}\hat{\mathbf{s}}_{\mathbf{k}}$ for $\hat{H}^a_0$,
therefore $\hat{H}_{I0}$ actually implies a direct coupling between the
electric current and the impurity spins, which provides a mechanism to
control the impurity spins through the electric voltage for dilute
impurity concentration.

\emph{RKKY interaction}: For simplicity we focus on the case of two
impurities at $\mathbf{r}_1$ and $\mathbf{r}_2$, respectively. The
results given below via the symmetry analysis are also valid for many
impurity case as long as we consider two-body interactions. The RKKY
interaction can be obtained by integrating out the degree of freedom of
conduction electrons, followed by an expansion to the second order of
the coupling parameters
$\lambda_{z,\pm}$\cite{ruderman1954,Kasuya1956,kei1957}. The effective
interaction between impurity spins thus obtained contains only the even
order terms for time reversal invariant system.

To the lowest order, the most general form of the interaction is
$\hat{H}_{rkky}(\mathbf{r}_{12}) = \sum_{\alpha_1\alpha_2}
\Gamma_{\alpha_1\alpha_2}(\mathbf{r}_{12}) \hat{S}^{\alpha_1}_1
\hat{S}^{\alpha_2}_2$ with $\mathbf{r}_{12}\equiv
\mathbf{r}_1-\mathbf{r}_2$. Without spin-orbit coupling as in the usual
case, only terms with $\alpha_1=\alpha_2$ can exist due to a rotational
symmetry in spin space.  However, the situation is quite different for
the helical metal, since $\hat{H}^a_{0}$ is invariant only under a
\emph{joint} rotation in both spin and orbital space, which allows more
terms as soon become clear. Let us consider a global rotation around $z$
axis by $\varphi$ with the form $\hat{\mathbf{R}}_{\mathbf{e}_z,\varphi}
\equiv \exp[ i \varphi \mathbf{e}_z \cdot \hat{\mathbf{s}}_{cond} ]
\times \exp[ i \varphi \mathbf{e}_{z} \cdot (\hat{\mathbf{S}}_1 +
\hat{\mathbf{S}}_2)] \times \exp[ i \varphi \mathbf{e}_{z} \cdot
\hat{\mathbf{L}}]$ with total spin of conduction electrons
$\hat{\mathbf{s}}_{cond}= \int d\mathbf{r}
\hat{c}^{\dagger}(\mathbf{r})\boldsymbol{\sigma} \hat{c}(\mathbf{r})/2$
and orbital angular momentum $\hat{\mathbf{L}}\equiv \int
d\mathbf{r}\hat{c}^{\dagger}(\mathbf{r})(\mathbf{r}\times \mathbf{p})
\hat{c}(\mathbf{r}) $.  Under this rotation, $\hat{H}^{a}_0$ is
obviously invariant, but
$\hat{H}_I(\mathbf{r}_1,\mathbf{r}_2)\rightarrow
\hat{H}_I(\mathbf{r}'_1,\mathbf{r}'_2)$ with $\mathbf{r}' \equiv
e^{-i\varphi(xp_y-yp_x)}\mathbf{r}e^{i\varphi(xp_y-yp_x)}$, which breaks
the rotational symmetry. However since the energy behaves like a scalar,
one expects $\hat{H}_{rkky}(\mathbf{r}_{12}) =
\hat{H}_{rkky}(\mathbf{r}'_{12})$, then what we are left with is to
construct $\hat{\mathbf{R}}_{\mathbf{e},\varphi}$-invariants with
vectors $\mathbf{e}_{12}(\equiv \mathbf{r}_{12}/r_{12})$,
$\hat{\mathbf{S}}_1$ and $\hat{\mathbf{S}}_2$. 

There are many such $\hat{\mathbf{R}}_{\mathbf{e}_z,\varphi}$-invariants
terms, most of which can be further eliminated by the following two
symmetries. The first one is exchanging $\mathbf{r}_1$ and
$\mathbf{r}_2$ which obviously leaves $\hat{H}$ invariant. Thus terms
like $\mathbf{e}_z\cdot(\hat{\mathbf{S}}_1\times \hat{\mathbf{S}}_2)$
are not allowed. The second one is a \emph{global} rotation
$\hat{\mathbf{R}}_{\mathbf{e}_{12},\pi}$ around axis $\mathbf{e}_{12}$
by $\pi$, which rules out invariants like $[\mathbf{e}_{12}\times
(\mathbf{S}_1 \times\mathbf{S}_2)]_z $ that changes sign under rotation
$\mathbf{R}_{\mathbf{e}_{12},\pi}$. Hence, we are left with the
following two terms in addition to the conventional ones,
$\mathbf{e}_{12}\cdot(\hat{\mathbf{S}}_1\times \hat{\mathbf{S}}_2)$ and
$(\mathbf{e}_{12}\cdot \hat{\mathbf{S}}_1)(\mathbf{e}_{12}\cdot
\hat{\mathbf{S}}_2)$. One then constructs the effective Hamiltonian of
impurity spins as
\begin{eqnarray}
\label{eq:3}
\hat{H}_{rkky}(\mathbf{r}_{12})&=&\Gamma_{0}(r_{12})[\lambda^2_{\pm} (\hat{S}_1^x\hat{S}^x_2 +
\hat{S}_1^y\hat{S}^y_2) + \lambda^2_z \hat{S}_1^z\hat{S}^z_2] \nonumber\\
&&+\Gamma_{1}(r_{12})\lambda_{\pm}^2[(\mathbf{e}_{12}\cdot\hat{\mathbf{S}}_1)(\mathbf{e}_{12}\cdot\hat{\mathbf{S}}_2)]
\nonumber\\ 
&&+\Gamma_{DM}(r_{12})\lambda_z\lambda_{\pm}[\mathbf{e}_{12}\cdot
(\hat{\mathbf{S}}_1\times \hat{\mathbf{S}}_2)],
\end{eqnarray}
where the $\Gamma_0$-term is just the conventional RKKY terms.

\emph{Derivation of $\Gamma_{0,1,DM}$}: The coefficients
$\Gamma_{0,1,DM}$ in Eq.~\eqref{eq:3} can be computed following RKKY's
second order perturbation\cite{ruderman1954,Kasuya1956,kei1957}.  Let us
consider the following generalized partition function which is partially
traced over conduction electrons, $\mathcal{Z} =
\text{Tr}_{\text{cond}}e^{-\beta(\hat{H}_0+\hat{H}_{I})} $.  To the
quadratic order of $\lambda_{z,\pm}$, we obtain
\begin{widetext}
\begin{eqnarray}
\label{eq:4}
\frac{\mathcal{Z}}{\mathcal{Z}_0}\approx\exp\left\{\frac{\beta }{\mathcal{V}^2}
\sum_{\mathbf{k}_1s_1}\sum_{\mathbf{k}_2s_2}\sum_{n,m}
\frac{f_{\mathbf{k}_1s_1}(1-f_{\mathbf{k}_2s_2})}{\xi_{\mathbf{k}_2s_2}-\xi_{\mathbf{k}_1s_1}} 
[\sum_{a=x,y,z}\lambda_{a}F^{a}_{\mathbf{k}_1s_1;\mathbf{k}_2s_2}
\hat{S}^{a}_{n}] [\sum_{b=x,y,z}\lambda_{b}F^{b}_{\mathbf{k}_2s_2;\mathbf{k}_1s_1}
\hat{S}^{b}_{m}] e^{i(\mathbf{r}_m-\mathbf{r}_n)(\mathbf{k}_1-\mathbf{k}_2)}\right\}
\end{eqnarray}
\end{widetext}
where $\mathcal{Z}_0$ is the partition function of conduction electrons,
$f_{\mathbf{k}s}=[1+e^{\beta(s\hbar |v_F|k-E_f)}]^{-1}$ is the Fermi
distribution function, and the three $F$-functions take the forms
$F^x_{\mathbf{k}_1s_1;\mathbf{k}_2s_2} = \frac{1}{4}
\text{sgn}(v_F)(s_1e^{-i\theta_{\mathbf{k}_1}}
+s_2e^{i\theta_{\mathbf{k}_2}})$, $F^y_{\mathbf{k}_1s_1;\mathbf{k}_2s_2}
=\frac{i}{4}\text{sgn}(v_F)(s_1e^{-i\theta_{\mathbf{k}_1}}
-s_2e^{i\theta_{\mathbf{k}_2}})$ and
$F^z_{\mathbf{k}_1s_1;\mathbf{k}_2s_2} =\frac{1}{4}(1-s_1 s_2
e^{-i(\theta_{\mathbf{k}_1}-\theta_{\mathbf{k}_2})})$, which depend only
on $\theta_{\mathbf{k}_{1,2}}$, the angle of $\mathbf{k}_{1,2}$. By
comparing with Eq.~\eqref{eq:3}, one can extract the coefficients
$\Gamma_{0,1,DM}$ from Eq.~\eqref{eq:4} straightforwardly. After
integrating over $\theta_{\mathbf{k}_{1,2}}$, we have
$\Gamma_{0}(r)=-2[A(r)-B(r)]$, $\Gamma_1(r)=-4B(r)$ and
$\Gamma_{DM}(r)=-2C(r)$, where the functions $A$, $B$ and $C$ read in
the zero temperature limit
\begin{eqnarray}
\label{eq:5}
A(r) &=&
\frac{1}{8\hbar |v_F|r^3}\sum_{s_1s_2}\int x_1dx_1
\int x_2dx_2
\mathscr{J}_0(x_1)\mathscr{J}_0(x_2)\nonumber\\
&&\hspace{1cm} \times\frac{\theta(x_F-x_1)\theta(x_2-x_{F})}{s_2x_2-s_1x_1} \nonumber\\
B(r) &=&
\frac{1}{8\hbar |v_F|r^3}\sum_{s_1s_2}\int x_1dx_1
\int x_2dx_2
\mathscr{J}_1(x_1)\mathscr{J}_1(x_2) \nonumber\\
&&\hspace{1cm}\times
\frac{\theta(x_F-x_1)\theta(x_2-x_{F})}{s_2x_2-s_1x_1} \nonumber\\
C(r)&=&\frac{\text{sgn}(v_F)}{8\hbar |v_F|r^3}\sum_{s_1s_2}\int x_1dx_1
\int x_2dx_2 \nonumber\\
&&\hspace{0.2cm}[
s_1\mathscr{J}_1(x_1)\mathscr{J}_0(x_2)-s_2\mathscr{J}_0(x_1)\mathscr{J}_1(x_2)]
 \nonumber\\
&&\times\frac{\theta(x_F-x_1)\theta(x_2-x_{F})}{s_2x_2-s_1x_1} 
\end{eqnarray}
where $x\equiv kr$, $x_F = k_Fr$, $\theta(x)$ is Heaviside step
function, and $\mathscr{J}_{0,1}(x)$ are the first and second order
Bessel functions, respectively. Note that $\Gamma_0(r)$ has already been
obtained in Ref.\cite{liuQ2009}, which agrees with ours. It is the
$\Gamma_{1}$ and $\Gamma_{DM}$ terms that have not been addressed
before, and the physics discussed in this letter is essentially coming
from them.

\begin{figure}[t]
\centerline{\includegraphics[width=4cm]{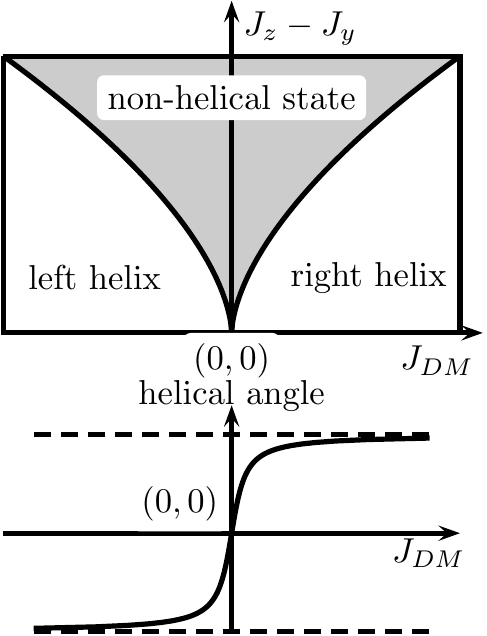}}
\caption[]{\label{fig:phasediagram} A schematic plot of phase diagram of
  $J_z-J_y$ vs. $J_{DM}$ in the upper panel for $K<1$, and helical angle
  as a function of $J_{DM}$ in lower panel. The left helix, non-helical
  state and right helix regions belong to different topological sectors
  of sine-Gordon model with negative, zero, and positive topological
  charges, respectively.}
\end{figure}

The numerical results of $\Gamma_{0,1,DM}(r)$ are plotted in
Fig.~\ref{fig:phasediagram} where a cutoff is set for $x$ and the Cauchy
principle integration is understood when singularities are encountered
in Eq.~\eqref{eq:5}.  Despite of the oscillation of the conventional
RKKY interaction with respect to $r$ for large $k_F$, it is interesting
to note that the DM interaction is of the same order of the conventional
ones, and depends on the sign of $v_F$ as seen from the expression of
$C(r)$ in Eq.~\eqref{eq:5}.

\begin{figure}[t]
\includegraphics[width=4cm]{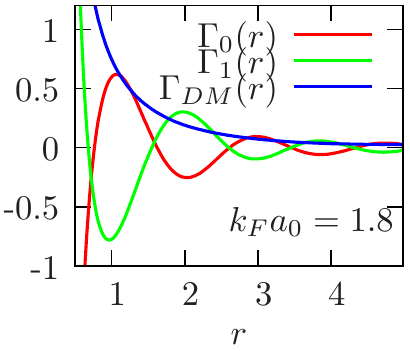}%
\includegraphics[width=4cm]{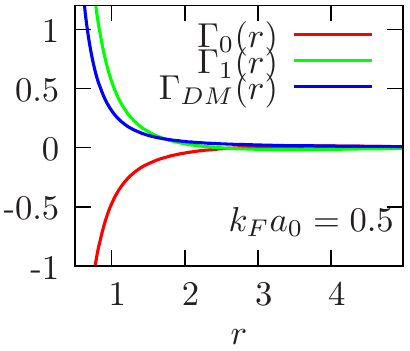}  
\caption[]{\label{fig:gamma}Plot of $\Gamma_{0,1,DM}$ as functions of
  the distance between impurities $r$, $k_{F}a_0=1.8$ (left) and
  $k_Fa_0=0.5$ (right). The distance between impurities is in unit of
  lattice spacing $a_0$. The coefficient $8\hbar v_{F}a_0^3$ is taken
  as one for convenience.}
\end{figure}

\emph{Single-handed Spin Helix:} To reveal the effect of DM interaction,
we focus on a simple case of one-dimensional chain of impurities aligned
with equal spacing $d$ along $x$-axis, and the corresponding Hamiltonian
reads
\begin{eqnarray}
\label{eq:6}
\hat{H}_{imp} &=& \sum_n
(-J_z\hat{S}^z_n\hat{S}^z_{n+1}-J_y\hat{S}^y_n\hat{S}^y_{n+1}
+J_x\hat{S}^x_n\hat{S}^x_{n+1}) \nonumber\\
&&+ J_{DM}(\hat{S}^z_n\hat{S}^y_{n+1}-\hat{S}^y_n\hat{S}^z_{n+1})
\end{eqnarray}
where $J_z=-\Gamma_0(d)\lambda^2_z$, $J_y=-\Gamma_0(d)\lambda^2_{\pm}$,
$J_x=[\Gamma_0(d)+\Gamma_1(d)]\lambda^2_{\pm}$, and
$J_{DM}=-\Gamma_{DM}(d)\lambda_z\lambda_{\pm}$. Here we consider the case
for $k_F$ close to the Dirac point, where we can simply assume
$J_{y,z}>J_x>0$ as can be read off in Fig.~\ref{fig:gamma}b for
$r\approx 1$. The other situations including two dimensional arrays of
magnetic impurities are left for future studies.

In this case, the spins are lying in $yz$ plane in the classical
limit. Therefore, we can introduce two variables $\hat{\theta}_n$ and
$\hat{\rho}_n$ to describe the spin rotation in $yz$-plane and the
amplitude fluctuation along $x$-axis, which satisfy the commutation
relation $[\hat{\theta}_n,\hat{\rho}_m] = i\delta_{m,n}$, and are
related to $\hat{\mathbf{S}}_n$ through $\hat{S}^x_n = \hat{\rho}$,
$\hat{S}^y_n + i \hat{S}^z_n=
e^{i\hat{\theta}}[(S-\hat{\rho})(S+\hat{\rho}+1)]^{1/2}$ and
$\hat{S}^y_n - i \hat{S}^z_n=
[(S-\hat{\rho})(S+\hat{\rho}+1)]^{1/2}e^{-i\hat{\theta}}$. One may
verify these equations satisfying the angular momentum algebra, and
$\hat{\mathbf{S}}^2=S(S+1)$. In continuum limit the effective
Hamiltonian for the dynamics of impurity spins can be written as
\begin{eqnarray}
\label{eq:7}
\hat{H}_{imp} &=& \int dx [ -D_1\cos(d\partial_x\hat{\theta}) +
D_2\cos(2\hat{\theta})\nonumber\\ 
&&+D_3\sin(d\partial_x\hat{\theta})+D_{0}\hat{\rho}^2] 
\end{eqnarray}
with $[\hat{\theta}(x),\hat{\rho}(x')]=i\delta(x-x')$. The parameters
are given by $D_0 = J_xd$, $D_1 = (J_z+J_{y})S(S+1)/(2d)$, $D_2 =
(J_z-J_{y})S(S+1)/(2d)$ and $D_{3}= J_{DM}S(S+1)/d$.

We first consider the isotropic case, i.e., $J_z=J_y$. The
$\hat{\theta}(x)$ field has a helical background(see
Fig.~\ref{fig:helix}(a)), with helical angle $\eta$ satisfying
$\tan(\eta d)=2J_{DM}/(J_z+J_y)$. The sign of $\eta$ is the same as that
of $J_{DM}$, and its amplitude is saturated to $\pi/2d$ as $J_{DM}$ is
large enough as shown in the lower panel in Fig.~\ref{fig:phasediagram}.
Replacing $\hat{\theta}(x)$ in Eq.~\eqref{eq:7} with $\eta x
+\hat{\theta}(x)$, we have $\hat{H}_{imp} = \int dx [
-\sqrt{D_1^2+D^2_{3}}\cos(d\partial_x\hat{\theta}) + D_{0}\hat{\rho}^2]
$ describing the quantum fluctuation upon the helical background. The
low energy excitations can be obtained by expanding
$\cos(d\partial_x\hat{\theta})$ to the second order of $d$, which leads
to a linear spectrum in $k$ as $\omega_{k} =
d(2D_0)^{1/2}(D_1^2+D_3^2)^{1/4} |k|$. 

Next we consider the effect of spin anisotropy, i.e. $J_z\ne
J_y$. Without loss of generality, we assume $J_z>J_y$(if $J_z<J_y$, one
can shift $\hat{\theta}\rightarrow \hat{\theta}+\pi/2$ to get positive
$D_2$). In this case the Hamiltonian becomes a sine-Gordon(SG) model
$\hat{H}_{imp}\approx \int dx
\frac{u}{2}[K^{-1}(\partial_x\hat{\theta})^2 +
K\hat{\rho}^2]+h\partial_x\hat{\theta}+D_2\cos(\sqrt{8\pi}\hat{\theta})$,
which is written in the standard form by rescaling $\hat{\theta}$ and
$\hat{\rho}$ in order to use the well known results in literature,
e.g. in Refs.\cite{gogolin1998,yang2001}. The coefficients take the form
$u=d \sqrt{J_x(J_y+J_z)S(S+1)}$, $K=\pi^{-1}\sqrt{J_x/(J_z+J_y)S(S+1)}$
and $h=\sqrt{2\pi}J_{DM}S(S+1)$.  In case $K>1$, the cosine term is
irrelevant, so that the system is massless and the spin helix takes
place for any finite $h$. However in case $K<1$, the theory turns out to
be massive, and a critical value of $|J_{DM}|$ exists, only above which
the spin helix can occur in forms of massive soliton excitations of
$\hat{\theta}$ field\cite{gogolin1998,yang2001} which connect different
classical vacua of the SG model.  In our case, since $0<J_x<J_{y,z}$, we
are in the region of $K<1$. A schematic phase diagram for this case is
given in the upper panel of Fig.~\ref{fig:phasediagram}, where the left
helix, non-helical and right helix regions correspond to different
topological sectors of sine-Gordon model with negative, zero, and
positive topological charges, respectively.

Our analysis of DM interaction so far is focused on $\hat{H}^a_0$, where
the spin helix is in $yz$ plane perpendicular to the impurity chain.
For helical metal described by $\hat{H}^b_0$, the RKKY interaction can
be obtained by rotating $\mathbf{e}_{12}$ around $z$-axis by $\pi/2$ in
Eq.~\eqref{eq:3}. As a consequence, the $\Gamma_0$-term is invariant,
$\Gamma_1$-term becomes
$\Gamma_1(r_{12})\lambda_{\pm}^2(\mathbf{e}_{12}\times
\hat{\mathbf{S}}_1)_z (\mathbf{e}_{12}\times \hat{\mathbf{S}}_2)_z$ and
$\Gamma_{DM}$-term becomes
$\Gamma_{DM}(r_{12})\lambda_{\pm}\lambda_{z}[\mathbf{e}_{12} \times
(\hat{\mathbf{S}}_1\times \hat{\mathbf{S}}_2)]_z$.  Notice that
$(\mathbf{e}_{12}\times \hat{\mathbf{S}}_1)_z (\mathbf{e}_{12}\times
\hat{\mathbf{S}}_2)_z = \hat{S}_1^x\hat{S}_2^x + \hat{S}_1^y
\hat{S}_2^y- (\mathbf{e}_{12}\cdot
\hat{\mathbf{S}}_1)(\mathbf{e}_{12}\cdot
\hat{\mathbf{S}}_2)$, therefore $\Gamma_1$-term simply changes the
coefficients of the first two terms of Eq.~\eqref{eq:3}. Only the change
of $\Gamma_{DM}$-term is essential, which makes the spin rotate in $xz$
plane instead of $yz$ plane as illustrated in Fig.~\ref{fig:helix}b.

\emph{Summary:} The surface state of TI is metallic with strong
spin-orbit coupling, in which the magnetic impurities coupled with
conductance electrons can be polarized by the electric voltage. There is
also an effective interaction between impurity spins mediated by the
conduction electrons, which includes a DM interaction with the same
order of amplitude of the isotropic RKKY interactions.  For 1D chain of
impurities, this could lead to a single-handed spin helix, and the
handedness is locked with the sign of Fermi velocity $v_F$ of the
emergent Dirac particles.

\acknowledgments{We thank C. X. Liu, D. Qian, J. Wang and Y. Y. Wang for
  stimulating discussions. FY is financially supported by NSFC Grant
  No. 10904081. HZ is supported by the Basic Research Young Scholars
  Program of Tsinghua University, NSFC Grant No. 10944002 and 10847002.}

%

\end{document}